\documentclass{article}

\usepackage{PRIMEarxiv}

\usepackage[utf8]{inputenc} 
\usepackage[T1]{fontenc}    
\usepackage{hyperref}       
\usepackage{url}            
\usepackage{booktabs}       
\usepackage{amsfonts}       
\usepackage{nicefrac}       
\usepackage{microtype}      
\usepackage{lipsum}
\usepackage{fancyhdr}       
\usepackage{graphicx}       
\graphicspath{{media/}}     
\usepackage{listings}
\usepackage{xcolor}

\title{DOCUEVAL: An LLM-based AI Engineering Tool for Building Customisable Document Evaluation Workflows}

\author{
  Hao Zhang, Qinghua Lu, Liming Zhu \\
  CSIRO's Data61 \\
  \texttt{\{firstname, lastname\}@data61.csiro.au} \\
}

\begin{document}
\maketitle

\begin{abstract}
Foundation models, such as large language models (LLMs), have the potential to streamline evaluation workflows and improve their performance. However, practical adoption faces challenges, such as customisability, accuracy, and scalability.
In this paper, we present \textbf{DOCUEVAL}, an AI engineering tool for building customisable \textbf{DOCU}ment \textbf{EVAL}uation workflows.
DOCUEVAL supports advanced document processing and customisable workflow design which allow users to define theory-grounded reviewer roles, specify evaluation criteria, experiment with different reasoning strategies and choose the assessment style. To ensure traceability, DOCUEVAL provides comprehensive logging of every run, along with source attribution and configuration management, allowing systematic comparison of results across alternative setups. By integrating these capabilities, DOCUEVAL directly addresses core software engineering challenges, including how to determine whether evaluators are “good enough” for deployment and how to empirically compare different evaluation strategies. We demonstrate the usefulness of DOCUEVAL through a real-world academic peer review case, showing how DOCUEVAL enables both the engineering of evaluators and scalable, reliable document evaluation.
\end{abstract}


\section{Introduction}

Document evaluation is a fundamental task across many professional domains, such as academic peer review and tender proposal evaluation. Foundation models~\cite{bass2025engineering}, such as large language models (LLMs), have the potential to streamline evaluation workflows and improve its efficiency, consistency, fairness, and scalability. However, applying LLM-based AI solutions to document evaluation in practice introduces significant software engineering challenges, such as customisation, accuracy, and scalability.

To address these challenges, in this paper, we present \textbf{DOCUEVAL}, an LLM-based AI engineering tool for building customisable \textbf{DOCU}ment \textbf{EVAL}uation workflows. DOCUEVAL supports advanced document processing and customisable workflow design which allow users to define theory-grounded reviewer roles rather than ad-hoc personas, specify evaluation criteria, experiment with alternative reasoning strategies (e.g., reasoning-before-scoring or reasoning-after-scoring), and choose the assessment style (e.g., scoring-based or qualitative analysis). To ensure reproducibility and traceability, DOCUEVAL provides comprehensive logging of every run, along with source attribution and configuration management, allowing systematic comparison of results across alternative setups. By integrating these capabilities, DOCUEVAL directly addresses core software engineering challenges, including how to determine whether evaluators are “good enough” for deployment and how to empirically compare different evaluation strategies.

The main contributions of our paper are: (1) A summary of common use cases and software engineering challenges in LLM-based AI document evaluation; (2) DOCUEVAL, an AI engineering tool for building customisable document evaluation workflows; (3) An evaluation using a real-world academic peer review use case, demonstrating the usefulness of DOCUEVAL; (4) A short demonstration video showcasing the key features of DOCUEVAL~\footnote{\url{https://drive.google.com/drive/folders/1-2nSiz0i0l-eixSnzk8bdIfXKXeEuQVT?usp=share_link}}.

The remainder of this paper is organised as follows. Section II summarises the use cases and challenges. Section III presents DOCUEVAL tool. Section IV evaluates the tool. Section V discusses related work. Section VI concludes the paper.

\section{Use Cases and Challenges in AI Document Evaluation}
Based on our project experience, we summarise four common uses cases for AI document evaluation.
One common scenario is \textit{criteria-aligned assessment}, where documents such as grant proposals must be evaluated against a set of well-defined criteria described in policies, rubrics, or guidelines. As workloads increase, AI systems can help by automatically screening, scoring, or classifying submissions based on these criteria. 
Another common use case is \textit{conformance checking}, in which a completed document is evaluated against a reference framework, such as a style manual, organisational policy, or predefined guardrails. 
A third scenario involves \textit{synthesis and thematic analysis}, where organisations must process large collections of source materials, such as public consultation submissions, research papers, or policy reports, and condense them into a single structured summary. The resulting report often serves as a critical input for decision-making, strategy development, or policy formulation.
Finally, many operational contexts require \textit{signal detection and prioritisation across multiple information streams}. Organisations must continuously scan incoming data, such as media reports, public advisories, or internal incident logs, to identify and prioritise emerging issues that demand timely attention.

We identify key software engineering challenges in building LLM-based document evaluations: (1) \textbf{Customisability:}  
    Off-the-shelf AI solutions may not align perfectly with specific evaluation workflows, criteria, reporting formats, etc. Building configurable and adaptable workflows is essential to support diverse evaluation needs across different domains. (2) \textbf{Accuracy:}  
    AI document evaluation tools often fail to achieve sufficient accuracy when dealing with complex or domain-specific documents. Subtle language, nuanced context, and specialised terminology can lead to misinterpretations and unreliable evaluation outcomes. Many documents contain non-textual elements such as images, tables, and diagrams which cannot be processed well. Current solutions often struggle to interpret these elements accurately, leading to incomplete or suboptimal evaluations. Internal policies may restrict the choice of LLMs with relatively small context windows, making it difficult to process documents effectively. (3) \textbf{Scalability:}  
    Organisations often require large-scale deployments capable of process vast numbers of documents efficiently. Scaling batch processing while maintaining low latency remains a persistent challenge. On the other hand, there are often large amounts of evaluation rules that evolve over time in response to new policies, regulations, or business requirements. These rules may be distributed across different documents and formats, making them difficult to track and manage. Maintaining, updating, and resolving conflicts among these rules introduces significant operational overhead and increases the risk of inconsistent evaluations.(4) \textbf{Privacy and Security:}  
    Many evaluation tasks involve sensitive or confidential documents. Using AI services hosted on public clouds or through external APIs introduces potential privacy and security risks.
     (5) \textbf{Meaningful human oversight:} While human-in-the-loop mechanisms are often included for review and validation, they are rarely designed in a systematical way to enable understandability and meaningful human oversight. Also, prompt versioning is ineffective, as the same prompt may not consistently produce identical outputs. Without comprehensive run-time logging and traceability, it becomes difficult to reproduce evaluation results or systematically compare alternative configurations.

\section{DOCUEVAL Tool}
\begin{figure}
    \centering
    \includegraphics[width=0.8\linewidth]{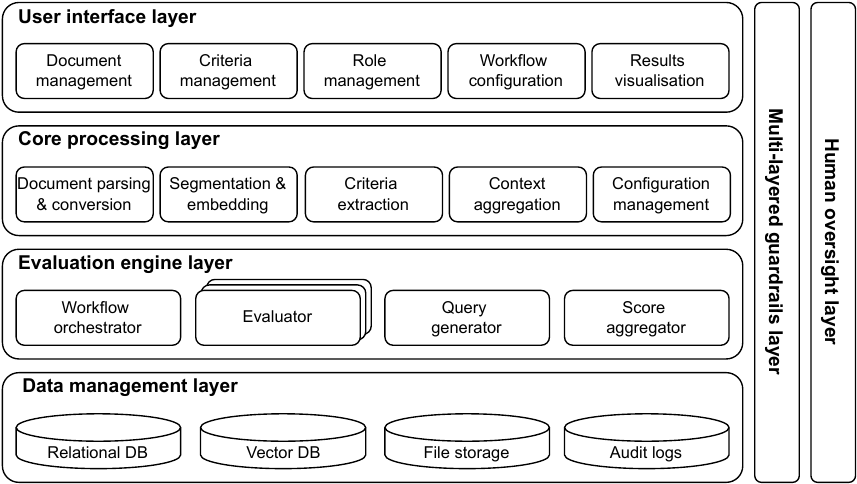}
    \caption{DOCUEVAL architecture.}
    \label{fig:architecture}
\end{figure}

In this section, we present the design of \textbf{DOCUEVAL}. 
As illustrated in Fig.~\ref{fig:architecture}, DOCUEVAL is organised into six key layers.

\subsection{User interface layer}
The user interface layer serves as the primary interaction point for users. Implemented as a React/Next.js web application, this layer allows users to upload, organise, and manage documents in multiple formats, including PDFs and Word files. Other than the document to evaluate, DOCUEVAL supports three types of support documents: criteria guidance that define assessment criteria, evaluation example that provide concrete examples, and reference standard that capture domain-specific knowledge.

This layer also allows users to customise evaluation workflows (customisability challenge), rather than relying on hard-coded, fixed pipelines. Users can define theory-grounded reviewer roles~\cite{manning2025general}, which go beyond simple persona-based approaches, to grounding these roles in established theories rather than ad hoc description. Rather than instructing the evaluator to “act like a harsh reviewer” or “be a supportive reviewer,” theory-grounded roles are built on formalised, well-defined theories. 
For example, in the context of academic peer review, the theory source includes document and studies on what is considered a high-quality review; the documents used by program committees to select the best reviewers for recognition or awards; a few slides and a few pages of documents prepared by professors on what is considered to be a good paper from a reviewer point of view; or empirical studies on what types of review is more useful or what characteristics define good papers.

Additionally, users can configure different reasoning strategies: (1) Reasoning-before-scoring generates a comprehensive rationale prior to assigning a score; (2) Reasoning-after-scoring produces the score first then provides an explanation to justify the assigned score; (3) Chain-of-thoughts generates a step-by-step reasoning path that lead to a final answer; (4) Deep-reasoning-with-planning performs in-depth reasoning and formulates a plan.

Evaluation criteria can be manually defined or automatically extracted from criteria guidance document uploaded, while assessment styles can be configured as either quantitative, score-based evaluations or qualitative, narrative reviews. Results are presented through interactive dashboards with clear visualisations, source attribution for every evaluation point, and comparison tools that allow users to analyse performance across different configurations and personas.

\subsection{Core processing layer}  
The core processing layer handles data processing and preparation. The document parsing and conversion service tackles the challenge of poor handling of non-textual elements by accurately reconstructing complex documents into formats that LLMs can understand. The workflow extracts text, figures, and tables using Optical Character Recognition (OCR) technology to preserve structural information, then converts them into markdown format while maintaining document hierarchies and relationships. Documents are intelligently segmented by the Segmentation and Embedding Service according to structural sections for retrieval-augmented generation (RAG), ensuring better retrieval accuracy by preserving contextual boundaries and logical document organisation. The architecture enables efficient batch processing capabilities, addressing the challenge of high document volumes (scalability challenge).

The criteria management service allows users to custom evaluation criteria tailored to their specific use cases. The service supports evaluator inheritance, allowing users to create new evaluators based on existing ones by copying established criteria sets (scalability challenge). Users can then add new criteria and update existing ones to create versioned evaluators that reflect evolving institutional requirements, regulatory changes, or updated assessment standards. This versioning system maintains backward compatibility while enabling seamless adaptation to changing evaluation needs.

The context aggregation service consolidates all relevant context including documents, criteria, and retrieved materials into structured inputs for evaluators.
The configuration management service handles configuration settings associated with each evaluator. User-specified configurations are captured and linked to the corresponding evaluator instance. These settings are also integrated with the workflow orchestrator, ensuring each evaluation run is executed consistently and fully aligned with the user-defined parameters.

\subsection{Evaluation engine layer}
The evaluation engine layer delivers evidence-based assessments by orchestrating configurable workflows that combine systematic reasoning with explicit evidence grounding. The workflow orchestrator service coordinates workflows by assessing each criterion individually. This approach mitigates context limitations (accuracy challenge) by reducing per-assessment context demands, enabling thorough evaluation of long or complex documents while maintaining consistency across criteria.
Each evaluator performs criterion-specific analysis using configurable roles and reasoning strategies. By applying different reasoning strategy, the evaluation process is broken into a series of steps. Each step draws on distinct pieces of evidence, and together these steps form a comprehensive body of evidence that supports the final evaluation result.
The dynamic query generator further improves accuracy by producing context-aware search queries tailored to criteria, document content, and supporting materials.
Finally, the score aggregator integrates individual criterion assessments into a coherent final evaluation, calculating an overall score based on user configurations, such as weighted average or simple average.

\subsection{Data management layer} 
The data management layer provides persistent storage and retrieval mechanisms for documents, metadata, and embeddings.
DOCUEVAL uses a multi-store approach, with each storage component optimised for specific data workloads.
A relational database manages structured information such as evaluator profiles, roles, evaluation criteria, and workflow configurations. A vector database stores embeddings generated from reference documents and evaluation examples, enabling high-performance semantic search and retrieval of relevant context during evaluation. A document repository manages both raw and processed files, preserving document versions and extracted components such as figures and diagrams. The audit logs record all human oversight activities, evaluation decisions, and system interactions for traceability.

\subsection{Guardrails layer}
The guardrails layer provides multiple, complementary built-in guardrails~\cite{shamsujjoha2025swiss} to prevent irresponsible or unreliable system behaviour (accuracy challenge, privacy and security challenge). At the user interface layer, guardrails validate inputs, ensuring that user entered data or uploaded documents are in the right format. Within the core processing layer, user inputs are validated to ensure prompts are free from potentially unethical or sensitive content such as personally identifiable information (PII) information. 
There are also guardrails monitoring document processing workflow integrity and addressing subtle prompt injection threats.
In the evaluation engine layer, guardrails focus on validating intermediate steps to detect hallucinations, factual inaccuracies, or unsupported claims, requiring all generated outputs to link back to verified sources.
Finally, at the data management layer, guardrails protect sensitive information through access control, encryption, and reference validation. 

\subsection{Human oversight layer}
The human oversight layer is designed to address the meaningful human oversight challenge and prevent over-reliance on AI by emphasising collaboration between humans and machines.
DOCUEVAL requires reviewers to complete their own independent evaluations before viewing AI-generated results. Once both assessments are completed, the system presents a side-by-side comparison, highlighting differences between human and AI evaluations and provide context-specific analysis.
Each AI generated result includes an explanation pack, containing a structured rationale table for every criterion: detailed justification, evidence excerpts from the source document, linked policy clauses or guideline references. This enables reviewers to verify claims without redoing the entire evaluation, focusing instead on whether the evidence truly supports the conclusion and whether the reasoning is consistent and traceable. 
Reviewers can annotate AI outputs, flag incorrect reasoning or results, and provide explanations. 
This feedback is logged and fed back into DOCUEVAL to improve future evaluations, enabling evaluation-driven learning.

\section{Evaluation}
\begin{figure}
  \centering
  \includegraphics[width=0.65\textwidth]{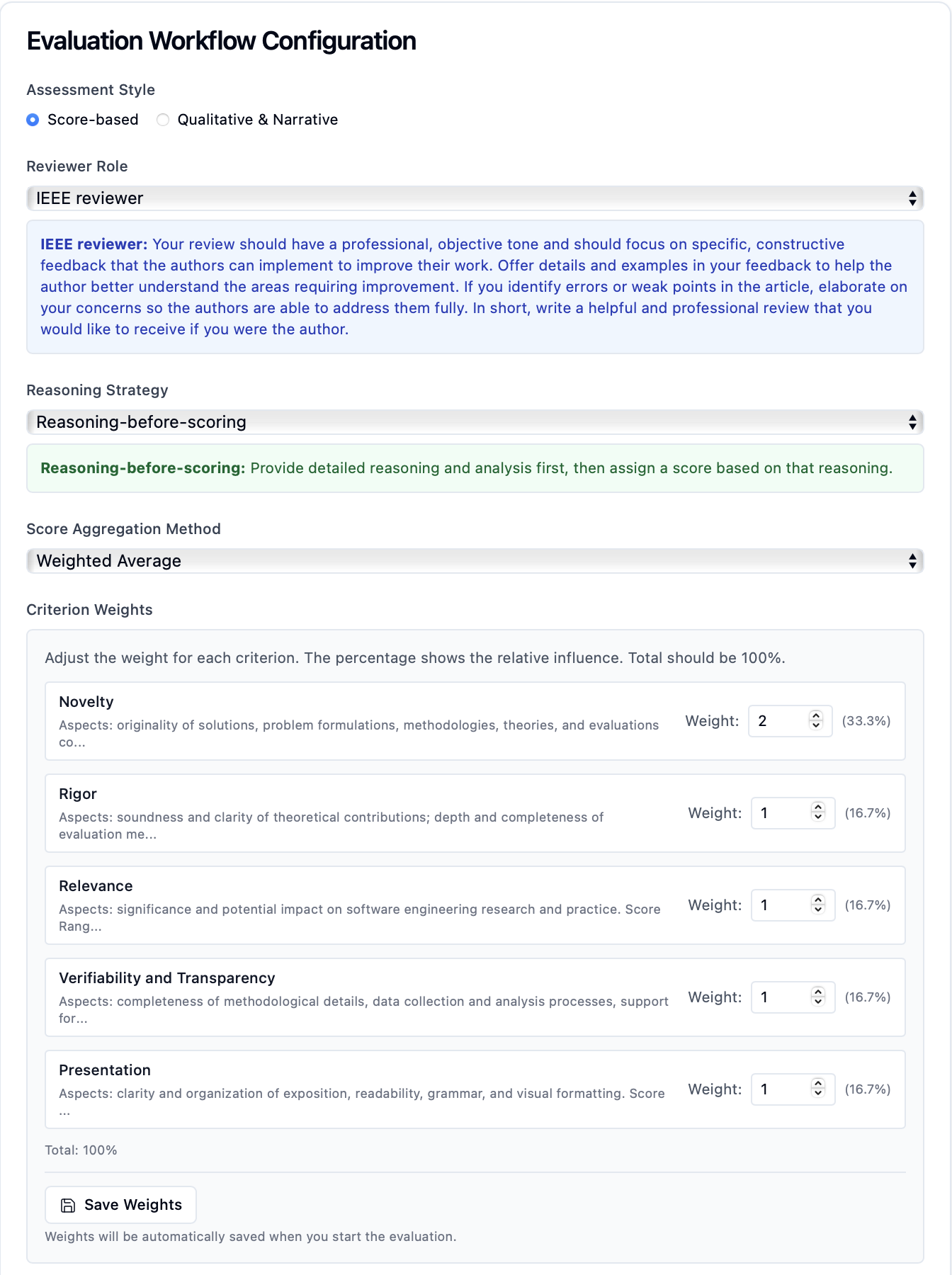}
  \caption{Evaluation workflow configuration screenshot}
  \label{fig:screenshot}
\end{figure}

In this section, we demonstrate the usefulness of DOCUEVAL through a real-world academic peer review use case. Specially, we simulate the process of reviewing a research paper, \textit{Agent Taxonomy}, submitted to the ICSE Research Track.
The process begins with the setup of the evaluator. The official ICSE'25 Research Track Review Criteria~\footnote{\url{https://conf.researchr.org/track/icse-2025/icse-2025-research-track?\#Call-for-Papers}} are used as criteria guidance, including novelty, rigor, relevance, verifiability and transparency, and presentation. The IEEE Conference Proceedings Formatting Guidelines are uploaded as reference standard. Once evaluator is set up, users could upload the paper, which is processed using the document parsing and conversion service, it preserves the document’s structure, tables, and figures to ensure accurate downstream evaluation. 
The tool also embedding relevant domain knowledge documents uploaded as reference standard, used for semantic search and RAG capabilities, ensuring that evaluations are grounded in domain-specific evidence. After that, users could start the configuration phase (Fig.~\ref{fig:screenshot}), where users build a tailored evaluation workflow rather than relying on a fixed pipeline.

We use ``Tips for Writing a Great Review'' in IEEE Reviewer Guidelines~\footnote{\url{https://ieeeauthorcenter.ieee.org/wp-content/uploads/ieee-reviewer-guidelines.pdf}} as the source for defining theory-grounded reviewer role.
Reasoning Before Score, which generates evidence and rationale before assigning numerical ratings, is selected as the reasoning strategies to determine how evaluations will be performed.

Scored-based evaluation style is selected, which assesses the paper numerically using weighted criteria, such as placing greater emphasis on novelty compared to other factors. 
The outputs are visualised through interactive dashboards, which include: (1) Source attribution: Every evaluation point is linked to specific passages in the paper or supporting documents, with clickable citations for full traceability; (2) Comparison of different evaluation workflow versions: Side-by-side visualisations of workflows to highlight how changes in role definitions, reasoning strategies, or criteria affect final outcomes; (3) Audit trails: Full logging of each evaluation run.

\section{Related Work}
Most LLM-based document evaluation tools provide limited customisation. For example, ReviewerGPT~\cite{liu2023reviewergpt} explored using large language models for academic peer review generation. However, their approach focused on generating review text rather than enabling configurable reviewer roles, alternative reasoning strategies, or domain-specific evaluation criteria.
Accuracy is core concerns in the literature. Several studies have investigated the reliability of LLM-generated evaluations. Efforts to improve accuracy often focus on RAG techniques~\cite{lewis2020rag}, but these approaches primarily address context limitations rather than end-to-end evaluation workflows. There is a need for integrated solutions that combine accuracy improvements with structured and transparent evaluation workflows. Handling complex and evolving rules for evaluation remains a significant research gap. Existing systems often rely on static rule sets extracted from regulations or business processes. For instance, ODRL-based pipelines allow formal compliance checking using translated legal rules \cite{devos2019odrl}, but they lack dynamic adaptation. 

Human oversight is critical for responsible AI but remain underdeveloped in the context of document evaluation. Research on human-in-the-loop (HITL) systems, such as Amershi et al. ~\cite{amershi2019guidelines}, demonstrates the benefits of hybrid human-AI decision-making. Recent studies of the UK government’s Consult platform~\cite{nesta-consult} have shown that structured human oversight significantly improves public trust and acceptance of AI-driven evaluations. Despite these insights, few existing systems systematically integrate oversight mechanisms into their design, leading to risks of automation bias and opaque decision-making.
While prior research has made progress on isolated aspects of document evaluation, gaps remain in customisation, accuracy, scalability, privacy, and human oversight. 

\section{Conclusion}

In this paper, we introduced DOCUEVAL, an AI engineering tool for customisable document evaluation workflows. Through a real-world academic peer review scenario, we demonstrated how DOCUEVAL enables flexible, versioned workflows, evidence-based assessments with full traceability, and effective collaboration between human reviewers and AI. Future work will focus on integrating more advanced evaluation-driven learning mechanisms to support the continuous improvement of evaluators.

\bibliographystyle{unsrt}  
\bibliography{references}

\end{document}